\documentclass[aps,showpacs,preprintnumbers]{revtex4}
\usepackage{amssymb}
\usepackage{amsmath}
\usepackage{bm}

\input{tcilatex}

\begin{document}

\title{Singularity-softening prescription for the Bethe-Salpeter equation.}
\author{J.H.O. Sales}
\affiliation{Instituto de Ci\^{e}ncias, Universidade Federal de Itajub\'{a}, CEP
37500-000, Itajub\'{a}, MG, Brazil}
\author{A.T. Suzuki}
\affiliation{Department of Physics, North Carolina State University, Raleigh, NC}
\date{\today }

\begin{abstract}
The reduction of the two fermion Bethe-Salpeter equation in the framework of
light-front dynamics is studied for one gauge $A^{+}=0$. The arising
effective interaction can be perturbatively expanded in powers of the
coupling constant $g$, allowing a defined number of gauge boson exchanges.
The singularity of the kernel of the integral equation at vanishs plus
momentum of the gauge is canceled exactly in on approuch. We studied the
problem using a singularity-softening prescription for the light-front gauge.
\end{abstract}

\pacs{12.39.Ki,14.40.Cs,13.40.Gp}
\maketitle

\address{ 27607.}









\section{Introduction}

The Bethe-Salpeter equation for ground state of two fermions exchanging
gauge boson presents divergences in the momentum $q^{+}$, even in the ladder
aproximation project in the light front \cite{jorgehenrique}. Gauge theories
with light front gauge, also present the difficulty associated to the
instantaneous term of the propagator of a system composed by fermions
boson-exchanger interaction. We used a prescription that allowed an
apropriate description of the singularity in the propagator of the gauge
boson in the light front.

We start by considering in Section $II$ gauge boson propagator is canonical
procedure of determining the propagator and projected in front of light,
that is, for intervals of time $x^{+}$. In section $III$ we obtain the two
free propagator only propagating modes are the physical. In Section $IV$ we
calculated the correction to the propagator of two fermions particles
spreading for the future in the light-front time change a gauge boson $g^{2}$%
. In Section $V$ we used the hierarchical equations to obtain Bethe-Salpeter
equation. In Section $VI$ we presented the in the light front Bethe-Salpeter
equation. In Section $VII$ we presented our conclusions.

\section{Gauge boson propagator}

The Lagrangian density for the vector gauge field (for simplicity we
consider an Abelian case) is given by 
\begin{equation}
\mathcal{L}=-\frac{1}{4}F_{\mu \nu }F^{\mu \nu }-\frac{1}{2\alpha }\left(
n_{\mu }A^{\mu }\right) ^{2}-\frac{1}{2\beta }\left( \partial _{\mu }A^{\mu
}\right) ^{2},  \label{lag}
\end{equation}%
where $\alpha $ and $\beta $ are arbitrary constants\cite{reex2}. Of course,
with these additional gauge breaking terms, the Lagrangian density is no
longer gauge invariant and as such gauge fixing problem in this sense do not
exist anymore. Now, $\partial _{\mu }A^{\mu }$ doesn't need to be zero so
that the Lorentz condition is verified \cite{Gupta}.

Here, instead of going through the canonical procedure of determining the
propagator as done in the previous section, we shall adopt a more head-on,
classical procedure by looking for the inverse operator corresponding to the
differential operator sandwiched between the vector potentials in the
Lagrangian density. For the Abelian gauge field Lagrangian density we have: 
\begin{equation}
\mathcal{L}=-\frac{1}{4}F_{\mu \nu }F^{\mu \nu }-\frac{1}{2\beta }\left(
\partial _{\mu }A^{\mu }\right) ^{2}-\frac{1}{2\alpha }\left( n_{\mu }A^{\mu
}\right) ^{2}=\mathcal{L}_{\text{E}}+\mathcal{L}_{GF}  \label{g1}
\end{equation}

By partial integration and considering that terms which bear a total
derivative don't contribute and that surface terms vanish since $%
\lim\limits_{x\rightarrow \infty }A^{\mu }(x)=0$, we have 
\begin{equation}
\mathcal{L}=\frac{1}{2}A^{\mu }\left( \square g_{\mu \nu }-\partial _{\mu
}\partial _{\nu }+\frac{1}{\beta }\partial _{\mu }\partial _{\nu }-\frac{1}{%
\alpha }n_{\mu }n_{\nu }\right) A^{\nu }  \label{g5}
\end{equation}

To find the gauge field propagator we need to find the inverse of the
operator between parenthesis in (\ref{g5}). That differential operator in
momentum space is given by: 
\begin{equation}
O_{\mu \nu }=-k^{2}g_{\mu \nu }+k_{\mu }k_{\nu }-\theta k_{\mu }k_{\nu
}-\lambda n_{\mu }n_{\nu }\,,  \label{g6}
\end{equation}
where $\theta =\beta ^{-1}$ and $\lambda =\alpha ^{-1}$, so that the
propagator of the field, which we call $S^{\mu \nu }(k)$, must satisfy the
following equation: 
\begin{equation}
O_{\mu \nu }S^{\nu \lambda }\left( k\right) =\delta _{\mu }^{\lambda }
\label{g7}
\end{equation}

$S^{\nu \lambda }(k)$ can now be constructed from the most general tensor
structure that can be defined, i.e., all the possible linear combinations of
the tensor elements that composes it \cite{reex1}: 
\begin{eqnarray}
S^{\mu \nu }(k) &=&g^{\mu \nu }A+k^{\mu }k^{\nu }B+k^{\mu }n^{\nu }C+n^{\mu
}k^{\nu }D+k^{\mu }m^{\nu }E+  \notag \\
&&+m^{\mu }k^{\nu }F+n^{\mu }n^{\nu }G+m^{\mu }m^{\nu }H+n^{\mu }m^{\nu
}I+m^{\mu }n^{\nu }J  \label{g8}
\end{eqnarray}%
where $m^{\mu }$ is the light-like vector dual ($m^{\mu }=n^{\ast \mu
}\equiv (n^{0},-\overrightarrow{n})$) to the $n^{\mu }$, and $A$, $B$, $C$, $%
D$, $E$, $F$, $G$, $H$, $I$ and $J$ are coefficients that must be determined
in such a way as to satisfy (\ref{g7}). Of course, it is immediately clear
that since (\ref{g5}) does not contain any external light-like vector $%
m_{\mu }$, the coefficients $E=F=H=I=J=0$ straightaway. We have 
\begin{eqnarray*}
S^{\mu \nu }(q) &=&-\frac{1}{q^{2}}\left\{ g^{\mu \nu }+\frac{\left( \alpha
q^{2}+n^{2}\right) \left( \beta -1\right) }{\left( \alpha q^{2}+n^{2}\right)
q^{2}+q^{+2}\left( \beta -1\right) }q^{\mu }q^{\nu }-\frac{q^{+}\left( \beta
-1\right) \left( q^{\mu }n^{\nu }+n^{\mu }q^{\nu }\right) }{\left( \alpha
q^{2}+n^{2}\right) q^{2}+q^{+2}\left( \beta -1\right) }+\right. \, \\
&&\left. -\left[ \frac{1}{\left( \alpha q^{2}+n^{2}\right)
q^{2}+q^{+2}\left( \beta -1\right) }\right] n^{\mu }n^{\nu }q^{2}\right\}
\end{eqnarray*}

Again, it is a matter of straightforward algebraic manipulation to get the
relevant propagator in the light-front gauge $n^{2}=0$ and taking the limit $%
\alpha $,$\beta \rightarrow 0$, we have 
\begin{equation}
S^{\mu \nu }(k)=-\frac{1}{q^{2}}\left\{ g^{\mu \nu }-\frac{q^{\mu }n^{\nu
}+n^{\mu }q^{\nu }}{q^{+}}+\frac{n^{\mu }n^{\nu }}{q^{+2}}q^{2}\right\} \,,
\label{g9}
\end{equation}
which has the outstanding third term commonly referred to as \emph{contact
term}. Now we needed a prescription to treat the singularity in $q^{+}$.

\section{Zero modes through the singularity-softening prescription}

Zero modes in the light front milieu is a very subtle problem which for
years have been challenging us with the best of our efforts to understand
it, to make it manageable and to make some physical sense out of it. We have
already learned that a prescription to handle those singularities cannot be
solely mathematically well-defined --- that is not enough --- we now know
that the prescription must be causal, that is, it needs to ascertain that
its implementation does not violate causality \cite{cp}. ML prescription has
been heralded as the causal prescription to handle the light-front
singularites and many a calculation do confirm that it has solved many
difficulties concerning one- and two-loop quantum corrections to Feynman
diagrams. However, as seen in the previous sections, ML prescription does
not remove the pathological zero modes in the one- and two-vector gauge
boson propagation at the quantum level. We therefore come to the place most
important in this work: The introduction of a novel prescription that is
causal and can handle the light-front singularities which does not leave
remnant zero modes is presented and applied to the one and two propagating
vector bosons in the light-front gauge.

The index \textrm{SS} stands for this singularity-softening prescription for
the treatment of the $(k^{+})^{-1}$ poles (cf.\cite{sing}), namely, 
\begin{eqnarray}
\left[ \frac{1}{k^{+}}\right] _{\mathrm{SS}} &=&\lim_{\varepsilon
\rightarrow 0}\left[ \frac{k^{2}}{k^{+}\left( k^{2}+i\varepsilon \right) }%
\right] _{\mathrm{SS}}  \notag  \label{ss} \\
&=&\lim_{\varepsilon \rightarrow 0}\left[ \frac{k^{-}-k_{\mathrm{on}}^{-}}{%
k^{+}\left( k^{-}-k_{\mathrm{on}}^{-}+\frac{i\varepsilon }{k^{+}}\right) }%
\right] _{\mathrm{SS}}
\end{eqnarray}

We use for one gauge boson. Like this, we will make a case for to component $%
S^{\perp -}$ is: 
\begin{eqnarray}
S^{(1)\perp -}(p^{-}) &=&i\int dk_{1}^{-}\frac{k_{1}^{\perp }\;n^{-}\delta
\left( p^{-}-k_{1}^{-}\right) }{k_{1}^{+}\left( k_{1}^{-}-\frac{k_{1}^{\perp
2}}{k_{1}^{+}}+\frac{i\varepsilon }{k_{1}^{+}}\right) }\left[ \frac{1}{%
k_{1}^{+}}\right] _{\mathrm{SS}}  \notag \\
&=&i\int dk_{1}^{-}\frac{k_{1}^{\perp }\;n^{-}\delta \left(
p^{-}-k_{1}^{-}\right) }{k_{1}^{+}\left( k_{1}^{-}-\frac{k_{1}^{\perp 2}}{%
k_{1}^{+}}+\frac{i\varepsilon }{k_{1}^{+}}\right) }\left[ \frac{%
k_{1}^{-}-k_{1\mathrm{on}}^{-}}{k_{1}^{+}\left( k_{1}^{-}-k_{1\mathrm{on}%
}^{-}+\frac{i\varepsilon }{k_{1}^{+}}\right) }\right] _{\mathrm{SS}}
\end{eqnarray}

The result is: 
\begin{equation}
S^{(1)\perp -}(p^{-})=\frac{\theta (p^{+})\;p^{\perp }n^{-}}{p^{+}}\left[ 
\frac{p^{-}-p_{\mathrm{on}}^{-}}{p^{+}\left( p^{-}-p_{\mathrm{on}}^{-}+\frac{
i\varepsilon }{p^{+}}\right) }\right] _{\mathrm{SS}}\frac{i}{\left(
p^{-}-K_{0}^{(1)-}+i\varepsilon \right) },
\end{equation}
where we have introduced the definition 
\begin{equation}
K_{0}^{(1)-}=p_{\mathrm{on}}^{-}=\frac{p^{\perp 2}}{p^{+}},
\end{equation}

However, since the Dirac delta funtion $\delta \left( p^{-}-k_{1}^{-}\right) 
$ forces us onto the mass-shell, the numerator is identically zero, that is, 
\begin{equation*}
p^{-}-p^-_{\mathrm{on}}=0
\end{equation*}
and this is true for massless as well as massive gauge bosons.

Thus, finally 
\begin{equation*}
S^{(1)\perp -}(p^{-})=0.
\end{equation*}

For the component $S^{--}$ we have $S^{(1)--}(p^{-})=0$.

In the case of $S^{(1)\perp \perp }(p^{-})$ component we have 
\begin{equation*}
S^{(1)\perp \perp }(P^{-})=\frac{\theta (p^{+})}{p^{+}}\frac{i\left(
-g^{\perp \perp }\right) }{\left( p^{-}-K_{0}^{(1)-}+i\varepsilon \right) }
\end{equation*}

Clearly, this case does not present us with the $p^{+}=0$ difficulty, and
the only non-vanishing result is just $S^{(1)\perp \perp }$. Only the
physical degrees of freedom (transverse ones) do propagate and without zero
mode hindrances anywhere!

The two gauge bosons case, $S^{(2)\perp -,\perp -}$ we have 
\begin{eqnarray}
S^{(2)\perp -,\perp -}(P^{-}) &=&-\frac{1}{\left( 2\pi \right) }\int \frac{%
dk_{1}^{-}}{k_{1}^{+}\left( P^{+}-k_{1}^{+}\right) }\left[ \frac{%
k_{1}^{-}-k_{1\mathrm{on}}^{-}}{k_{1}^{+}\left( k_{1}^{-}-k_{1\mathrm{on}%
}^{-}+\frac{i\varepsilon }{k_{1}^{+}}\right) }\right] _{\mathrm{SS}}\frac{%
n^{-}k_{1}^{\perp }}{\left( k_{1}^{-}-\frac{k_{1}^{\perp 2}}{k_{1}^{+}}+%
\frac{i\varepsilon }{k_{1}^{+}}\right) }  \notag \\
&\times &\left[ \frac{P^{-}-k_{1}^{-}-k_{2\mathrm{on}}^{-}}{\left(
P^{+}-k_{1}^{+}\right) \left( P^{-}-k_{1}^{-}-k_{2\mathrm{on}}^{-}+\frac{%
i\varepsilon }{P^{+}-k_{1}^{+}}\right) }\right] _{\mathrm{SS}}\frac{%
n^{-}k_{2}^{\perp }}{\left( P^{-}-k_{1}^{-}-\frac{\left( P^{\perp
}-k_{1}^{\perp }\right) ^{2}}{P^{+}-k_{1}^{+}}+\frac{i\varepsilon }{%
P^{+}-k_{1}^{+}}\right) }\ .  \notag
\end{eqnarray}

Evaluating the residue at the pole 
\begin{equation}  \label{pole}
k_{1}^{-}=k_{1\mathrm{on}}^{-}-\frac{i\varepsilon }{k_{1}^{+}}\text{ ,}
\end{equation}
we have 
\begin{equation*}
S^{(2)\perp -,\perp -}(P^{-})=0
\end{equation*}

For the other component we have

\begin{equation*}
\begin{array}{cc}
S^{(2)--,--}(P^{-})=0 & S^{(2)\perp -,--}(P^{-})=0 \\ 
S^{(2)\perp -,\perp \perp }(P^{-})=0 & S^{(2)--,\perp \perp }(P^{-})=0%
\end{array}%
\end{equation*}

Finally, for the component $S^{(2)\perp \perp ,\perp \perp }$ result 
\begin{equation}
S^{(2)\perp \perp ,\perp \perp }(P^{-})=-\frac{1}{\left( 2\pi \right) }\int 
\frac{dk_{1}^{-}}{k_{1}^{+}\left( P^{+}-k_{1}^{+}\right) }\frac{\left(
-g^{\perp \perp }\right) }{\left( k_{1}^{-}-\frac{k_{1}^{\perp 2}}{k_{1}^{+}}%
+\frac{i\varepsilon }{k_{1}^{+}}\right) }\frac{\left( -g^{\perp \perp
}\right) }{\left( P^{-}-k_{1}^{-}-\frac{\left( P^{\perp }-k_{1}^{\perp
}\right) ^{2}}{P^{+}-k_{1}^{+}}+\frac{i\varepsilon }{P^{+}-k_{1}^{+}}\right) 
}\ .  \notag
\end{equation}%
which yields the non-vanishing contribution 
\begin{equation*}
S^{(2)\perp \perp ,\perp \perp }(P^{-})=\frac{\theta (k_{1}^{+})\theta
(P^{+}-k_{1}^{+})}{k_{1}^{+}\left( P^{+}-k_{1}^{+}\right) }\frac{i\left(
-g^{\perp \perp }\right) \left( -g^{\perp \perp }\right) }{\left(
P^{-}-K_{0}^{(2)-}+i\varepsilon \right) },
\end{equation*}%
which is the same as that obtained through the ML-prescription \cite{ml}.

Therefore, as long as we treat the troublesome zero modes $k^{+}=0$ via the
singularity-softening prescription (\ref{ss}) the only non-vanishing
component of the two gauge boson propagator is the {\footnotesize $(\perp
\!\perp ,\perp \!\perp )$}-component, so that there is no zero mode problem
left and the only propagating modes are the physical, transverse ones!

\section{Green's Function $\mathcal{O}(g^{2})$}

To proceed we calculated the correction to the propagator of two particles
fermionic spreading for the future in the time in front of light changing a
gauge boson, in which the density Lagrangeana is given for 
\begin{equation}
\mathcal{L}_{I}=g\overline{\Psi }_{1}\gamma _{\mu }A^{\mu }\Psi _{1}+g%
\overline{\Psi }_{2}\gamma _{\nu }A^{\nu }\Psi _{2}.  \label{i0}
\end{equation}

The fields $\Psi _{1}$ and $\Psi _{2}$ they correspond to the fermions with
mass $m_{1}$ and $m_{2}$, that we will assume same $m_{1}=m_{2}=m$, and the
field $A^{\mu }$ it corresponds to the boson of intermediate gauge of null
mass. The coupling constant is $g$.

The correction to the propagator of two fermions corresponding to a diagram
of the type ladder, it is made calculations, in order $g^{2}$. For that we
will make the use just of the term propagate of the propagators fermionic of
the external lines.

The perturbative correction to the two-body propagator which comes from the
exchange of one intermediate virtual boson, is given by 
\begin{equation}
\Delta S_{g^{2}}(x^{+})=-\left( ig\right) ^{2}\int d\overline{x}_{1}^{+}d%
\overline{x}_{2}^{+}S_{k^{\prime }}(x^{+}-\overline{x}_{1}^{+})\gamma _{\mu
}^{(1)}S_{k}(\overline{x}_{1}^{+})S_{q}^{\mu \nu }(\overline{x}_{1}^{+}-%
\overline{x}_{2}^{+})S_{p^{\prime }}(x^{+}-\overline{x}_{2}^{+})\gamma _{\nu
}^{(2)}S_{p}(\overline{x}_{2}^{+}).  \label{i1}
\end{equation}%
The intermediate gauge boson, $q$, propagates between the time interval $%
\overline{x}_{1}^{+}-\overline{x}_{2}^{+}$. The labels in the particle
propagators $p$ and $k$ indicates initial and $p^{\prime }$ and $k^{\prime }$
final states. Propagator in the light-front gauge is 
\begin{equation}
S_{q}^{\mu \nu }(\overline{x}_{1}^{+}-\overline{x}_{2}^{+})=\frac{1}{2}\int 
\frac{dq^{-}}{2\pi }\frac{iN^{\mu \nu }(q)e^{-\frac{i}{2}\left( \overline{x}%
_{1}^{+}-\overline{x}_{2}^{+}\right) }}{(q^{+})^{3}\left( q^{-}-\frac{%
q^{\perp 2}}{q^{+}}+\frac{i\varepsilon }{q^{+}}\right) },  \label{i2}
\end{equation}%
where 
\begin{equation}
N^{\mu \nu }=-q^{+2}g^{\mu \nu }+q^{+}(q^{\mu }n^{\nu }+n^{\mu }q^{\nu
})-n^{\mu }n^{\nu }q^{2},  \label{i3}
\end{equation}%
\ $q^{+}=k^{\prime +}-k^{+}$ and $n^{\mu }=(0,2,0^{\perp })$.

This derivation is quite straightforward. However, the resulting equation
does not make sense as it stands. The most pressing problem is that there
are infrared divergences of the form $\int \frac{dq^{+}}{q^{+}}=\ln \left(
\infty \right) $.

The 1-body Green's functions can be derived from the covariant propagator
for 1-particles propagating at equal light-front times. In this case the
propagator from $x^{+}=0$ to $x^{+}>0$ is given by

\begin{equation}
S(x^{+})=\frac{1}{2}\int \frac{dk^{-}dk^{+}dk^{\perp }}{\left( 2\pi \right) }%
\frac{ie^{\frac{-i}{2}k^{-}x^{+}}}{k^{+}\left( k^{-}-\frac{k_{\perp
}^{2}+m^{2}}{k^{+}}+\frac{i\varepsilon }{k^{+}}\right) }.  \label{2b}
\end{equation}

The Fourier transform to the total light-front energy $(P^{-})$ is given by $%
{S}(P^{-})=\frac{1}{2}\int dx^{+}e^{\frac{i}{2}P^{-}x^{+}}S(x^{+})$ and the
free 1-body Green's function is given by $S(k^{-})=\frac{1}{k^{+}}G(k^{-})$,
where$\ $%
\begin{equation}
G_{0}^{(1)}(k^{-})=\frac{\theta (k^{+})}{k^{-}-k_{on}^{-}}\   \label{4b}
\end{equation}
with $k_{on}^{-}=\frac{k_{\perp }^{2}+m^{2}}{k^{+}}$ being the light-front
Hamiltonian of the free 1-particle system.

Let $S_{\text{F}}$ denote fermion field propagator in covariant theory 
\begin{equation}
S_{\text{F}}(x^{\mu })=\int \frac{d^{4}k}{\left( 2\pi \right) ^{4}}\frac{i(%
\rlap\slash k_{\text{on}}+m)}{k^{2}-m^{2}+i\varepsilon }e^{-ik^{\mu }x_{\mu
}},  \label{4}
\end{equation}
where $\rlap\slash k_{\text{on}}=\frac{1}{2}\gamma ^{+}\frac{(k^{\perp
})^{2}+m^{2}}{k^{+2}}+\frac{1}{2}\gamma ^{-}k^{+}-\gamma ^{\perp }k^{\perp }$%
. Using light-front variables in the Eq.(\ref{4}), we have 
\begin{equation}
S_{\text{F}}(x^{+})=\frac{i}{2}\int \frac{dk^{-}dk^{+}dk^{\perp }}{\left(
2\pi \right) }\left[ \frac{\rlap\slash k_{on}+m}{k^{+}\left(
k^{-}-k_{on}^{-}+\frac{i\varepsilon }{k^{+}}\right) }+\frac{\gamma ^{+}}{%
2k^{+}}\right] e^{\frac{-i}{2}k^{-}x^{+}}.  \label{5}
\end{equation}

We note that for the fermion field, light-front propagator differs from the
Feynmam propagator by an instantaneous propagator.

The free 1-fermion Green's function is given by 
\begin{equation}
G(k^{-})=\frac{\Lambda _{+}\left( k\right) }{\left( k^{-}-k_{on}^{-}+\frac{%
i\varepsilon }{k^{+}}\right) },  \label{7.21}
\end{equation}
where $\Lambda _{\pm }\left( k\right) =\frac{\pm \rlap\slash k_{on}+m}{2m}%
\theta (\pm k^{+})$.

\section{Coupled equations for the Green's functions}

Using the technique of the hierarchical equations \cite{hieraquia} together
with the paper of the reference \cite{21}, \ we built in the light-front
Green's function for the two fermions system obtained from the solution of
the covariant BS equation, that contains all two-body irreducible diagrams,
with the exception of those including closed loops of bosons $\Psi _{1}$ and 
$\Psi _{2}$ and part of the cross-ladder diagrams, is given by: 
\begin{equation}
G^{(2)}(K^{-})=G_{0}^{(2)}(K^{-})+G_{0}^{(2)}(K^{-})VG^{(3)}(K^{-})VG^{(2)}(K^{-})\ ,
\label{3.28}
\end{equation}%
\begin{equation}
G^{(3)}(K^{-})=G_{0}^{(3)}(K^{-})+G_{0}^{(3)}(K^{-})VG_{0}^{(4)}(K^{-})VG^{(3)}(K^{-}).
\label{3.29}
\end{equation}%
In the Yukawa model for fermions, the interaction operator acting between
Fock-states differing by zero, one and two $\sigma $'s, has matrix elements
given by 
\begin{eqnarray}
&&\langle (q,s^{\prime })k_{\sigma }|V|(k,s)\rangle =-2m(2\pi )^{3}\delta
(q+k_{\sigma }-k)\frac{g_{S}}{\sqrt{q^{+}k_{\sigma }^{+}k^{+}}}\theta
(k_{\sigma }^{+}){\overline{u}}(q,s^{\prime })u(k,s)~,  \label{mia} \\
&&\langle (q,s^{\prime })k_{\sigma }^{\prime }|V|(k,s)k_{\sigma }\rangle
=-2(2\pi )^{3}\delta (q+k_{\sigma }^{\prime }-k-k_{\sigma })\delta
_{s^{\prime }s}\frac{g_{S}^{2}}{\sqrt{k_{\sigma }^{\prime +}k_{\sigma }^{+}}}%
{\frac{\theta (k_{\sigma }^{\prime +})\theta (k_{\sigma }^{+})}{%
k^{+}+k_{\sigma }^{+}}}~,  \label{mib} \\
&&\langle (q,s^{\prime })k_{\sigma }^{\prime }k_{\sigma }|V|(k,s)\rangle
=-2(2\pi )^{3}\delta (q+k_{\sigma }^{\prime }+k_{\sigma }-k)\delta
_{s^{\prime }s}\frac{g_{S}^{2}}{\sqrt{k_{\sigma }^{\prime +}k_{\sigma }^{+}}}%
{\frac{\theta (k_{\sigma }^{\prime +})\theta (k_{\sigma }^{+})}{%
k^{+}-k_{\sigma }^{+}}}~.  \label{mic}
\end{eqnarray}%
The instantaneous terms in the two-fermion propagator give origin to Eqs. (%
\ref{mib}) and (\ref{mic}).

A systematic expansion by the consistent truncation of the light-front Fock
space up to $N$ particles in the intermediate states (boson 1, boson 2 and $%
N-2$ $\sigma $'s) in the set of Eqs.(\ref{3.29}) and (\ref{3.28}), amounts
to substitution $G^{(3)}(K^{-})\cong G_{0}^{(3)}(K^{-})$. The kernel of Eq.(%
\ref{3.28}) still contains an infinite sum of light-front diagrams, that are
obtained solving by Eq.(\ref{3.29}). To obtain the ladder aproximation up to
order $g^{2}$, Eq.(\ref{3.28}), only the free and first order terms are kept
in Eq.(\ref{3.29}), with the restriction of only one and one boson covariant
exchanges.Therefore, we have for Eq.(\ref{3.28}) 
\begin{equation}
G_{g^{2}}^{(2)}(K^{-})=G_{0}^{(2)}(K^{-})+G_{0}^{(2)}(K^{-})VG_{0}^{(3)}(K^{-})VG_{g^{2}}^{(2)}(K^{-}),
\label{5.1a}
\end{equation}
Taking the two-boson system as an example and restricting the intermediate
state propagation up to 3-particles, we find that 
\begin{equation}
G_{g^{2}}^{(2)}(K^{-})=G_{0}^{(2)}(K^{-})+G_{0}^{(2)}(K^{-})VG_{0}^{(3)}(K^{-})V\left\{ G_{0}^{(2)}(K^{-})+G_{0}^{(2)}(K^{-})VG_{0}^{(3)}(K^{-})VG_{g^{2}}^{(2)}(K^{-})\right\}
\label{3.29a}
\end{equation}
The correction in order $g^{2}$ is given for $\Delta
G_{g^{2}}^{(2)}(K^{-})=G_{0}^{(2)}(K^{-})VG_{0}^{(3)}(K^{-})VG_{0}^{(2)}(K^{-}) 
$.

\section{Bethe-Salpeter equation}

We perform the quasi-potential reduction of two-body BSE's and present the
coupled set of equations for the light-front Green's functions for gauge
boson and fermion models, with the interaction Lagrangian respectively given
by (\ref{i0}).

The bound state the Green's function has a pole $\lim_{K^{-}\rightarrow
K_{B}^{-}}G^{\left( 2\right) }(K^{-})=\frac{\left\vert \psi
_{B}\right\rangle \left\langle \psi _{B}\right\vert }{K^{-}-K_{B}^{-}}$,
where $\left\vert \psi _{B}\right\rangle $ it is the wave-function of the
bound state. Therefore, the homogeneous equation for the light-front
two-body bound state wave-function is obtained the solution of 
\begin{equation}
|\Psi _{B}>=G_{0}^{(2)}(K_{B}^{-})VG^{(3)}(K_{B}^{-})V|\Psi _{B}>,
\label{5.3}
\end{equation}%
the vertex function for the bound state wave-function is defined as 
\begin{equation}
\Gamma _{B}(k_{\perp },q^{+})=<k,\ K-k|\left( G_{0}^{(2)}(K_{B}^{-})\right)
^{-1}|\Psi _{B}>.  \label{vertexlf}
\end{equation}

The Green's function obtained from this equation, up to order $g^{2}$,
reproduces the covariant two-body propagator between two light-front
hypersurfaces. In this approximation, the vertex function satisfies the
following integral equation, 
\begin{equation}
\Gamma _{B}(\overrightarrow{q}_{\perp },y)=\int \frac{dxd^{2}k_{\perp }}{%
x(1-x)}\frac{\mathit{K}^{(3)}(\overrightarrow{q}_{\perp },y;\overrightarrow{k%
}_{\perp },x)}{M_{B}^{2}-M_{0}^{2}}\Gamma _{B}(\overrightarrow{k}_{\perp
},x),  \label{5.7}
\end{equation}%
where the momentum fractions are $y=q^{+}/K^{+}$ and $x=k^{+}/K^{+}$, with $%
0<y<1$. Where $\overrightarrow{K}_{\perp }=0$, and $%
M_{0}^{2}=K^{+}K_{(2)on}^{-}-K_{\perp }^{2}=\frac{k_{\perp }^{2}+m^{2}}{%
x(1-x)}$. The part of the kernel which contains only the propagation of
virtual three particle states foward in the light-front time is obtained
from Eq.(\ref{5.7}) as, 
\begin{eqnarray}
\mathcal{K}^{(3)}(y,q_{\perp };x,k_{\perp }) &=&\frac{g^{2}}{16\pi ^{3}}%
\frac{\Lambda _{+(1)}(q)\Lambda _{+(1)}(k)\Lambda _{+(2)}(K-q)\Lambda
_{+(2)}(K-k)\theta (y-x)}{\left( y-x\right) ^{2}\left( M_{B}^{2}-\frac{%
\overrightarrow{q}_{\perp }^{2}+m^{2}}{1-y}-\frac{\vec{k}_{\perp }^{2}+m^{2}%
}{x}-\frac{(\overrightarrow{q}_{\perp }-\vec{k}_{\perp })^{2}+\mu ^{2}}{y-x}%
\right) }+  \notag \\
&&+\left[ k\leftrightarrow q\right] ,  \notag
\end{eqnarray}%
being $M_{B}^{2}=K_{B}^{+}K_{B}^{-}$. We got the attention for the
relationship between $\left\vert \Gamma _{B}\right\rangle $ and $\Gamma _{B}(%
\overrightarrow{q}_{\perp },y)$, defined for $\Gamma _{B}(\overrightarrow{q}%
_{\perp },y)=\sqrt{q^{+}(K^{+}-q^{+})}\left\langle \overrightarrow{q}_{\perp
},q^{+}\right. \left\vert \Gamma _{B}\right\rangle $.

\section{Conclusion}

Bethe-Salpeter equation (\ref{5.7}) presents divergence in his kernel. We
hoped to solve the problem of singularity that appears in the equation of
Bethe-Salpeter with the inclusion of the singularity-softening prescription.
Like this, to obtain finite Bethe-Salpeter. We are still accomplishing
calculations for that aim at.

\textit{Acknowledgments:} J.H.O.Sales thanks the hospitality of Instituto de
F\'{\i}sica Te\'{o}rica/UNESP, where part of this work has been
done.A.T.Suzuki thanks the kind hospitality of Physics Department, North
Carolina Staste University and gratefully acknowledges partial support from
CNPq (Bras\'{\i}lia) in the earlier stages of this work, now superseded by a
grant from CAPES (Bras\'{\i}lia).


\begin{thebibliography}{9}
\bibitem{jorgehenrique} J.H.O.Sales, Tobias Frederico and B.M.Pimentel,
Hadron Physiscs, 211-214 (2002).

\bibitem{reex2} A.T. Suzuki and J.H.O.Sales, Mod. Phys. Lett. \textbf{A19 }%
1925 (2004).

\bibitem{Gupta} C.Itzykson and J.-B.Zuber, \ \textquotedblleft Quantum field
theory \textquotedblright , McGraw-Hill, 1980.

\bibitem{reex1} A.T. Suzuki and J.H.O.Sales, Nucl.Phys. \textbf{A725} (2003)
139.

\bibitem{cp} B.M.Pimentel and A.T.Suzuki, Phys.Rev.\textbf{D42} , 2115
(1990).

\bibitem{sing} A.T. Suzuki and J.H.O. Sales, Mod. Phys. Lett. \textbf{A19}
2831 (2004).

\bibitem{ml} S.Mandelstam, Nucl.Phys. \textbf{B213}, 149 (1983). G.
Leibbrandt, Phys. Rev. \textbf{D29}, 1699 (1984).

\bibitem{hieraquia} T. Frederico, J.H.O. Sales, B.V. Carlson e P.U. Sauer,
Nucl.Phys. \textbf{A737} (2004) 260.

\bibitem{21} J.H.O.Sales, T. Frederico, B.V. Carlson e P.U. Sauer, Phys.
Rev. \textbf{C} \textbf{61} (2000) 044003.
\end{thebibliography}
\end{document}